\documentclass[twocolumn,english,aps,prl,showpacs,superscriptaddress,amsmath,amssymb]{revtex4}
\usepackage[T1]{fontenc}
\usepackage[latin9]{inputenc}
\usepackage{textcomp}
\usepackage{amstext}
\usepackage{graphicx}

\makeatletter

\newcommand{\lyxmathsym}[1]{\ifmmode\begingroup\def\b@ld{bold}
  \text{\ifx\math@version\b@ld\bfseries\fi#1}\endgroup\else#1\fi}

\newcommand{\lyxdot}{.}

\@ifundefined{textcolor}{}
{%
 \definecolor{BLACK}{gray}{0}
 \definecolor{WHITE}{gray}{1}
 \definecolor{RED}{rgb}{1,0,0}
 \definecolor{GREEN}{rgb}{0,1,0}
 \definecolor{BLUE}{rgb}{0,0,1}
 \definecolor{CYAN}{cmyk}{1,0,0,0}
 \definecolor{MAGENTA}{cmyk}{0,1,0,0}
 \definecolor{YELLOW}{cmyk}{0,0,1,0}
 }

\usepackage{epsfig}
\usepackage{bm}
\usepackage{bbm}

\usepackage{epic}
\usepackage{eepic}
\usepackage{pifont}
\@ifundefined{definecolor}
 {\usepackage{color}}{}

\usepackage{nicefrac}
\input epsf
\input rotate

\newcommand{\be}{\begin{equation}}
\newcommand{\ee}{\end{equation}}
\newcommand{\bea}{\begin{eqnarray}}
\newcommand{\eea}{\end{eqnarray}}

\newcommand{\comment}[1]{}

\makeatother

\usepackage{babel}

\begin{document}

\title{Numerical study of 3D vesicles under flow: discovery of new peculiar behaviors}

\author{Thierry Biben}

\affiliation{{ Universit\'e de Lyon, F-69000, France; Univ. Lyon 1, Laboratoire PMCN; CNRS, UMR 5586; F-69622 Villeurbanne Cedex}}

\author{Alexander Farutin} \author {Chaouqi Misbah}

\affiliation{Laboratoire de Spectrom\'etrie Physique, UMR, 140 avenue de la physique,
Universit\'e Joseph Fourier Grenoble, and CNRS, 38402 Saint Martin d'H\`eres,
France 
Date: \today}

\begin{abstract}
The study of vesicles under flow, a model system for red blood cells (RBCs), is an essential step in understanding various intricate dynamics
exhibited by RBCs in vivo and in vitro.
Quantitative 3D analyses of vesicles under flow are presented.  The regions of parameters to produce tumbling (TB),
tank-treating (TT), vacillating-breathing (VB) {and even Kayaking (K)} modes are determined. New qualitative features are found:  (i) a significant
widening of the VB mode region in parameter space upon increasing
shear rate $\dot\gamma$ and (ii) a striking robustness of period of TB and VB
with $\dot\gamma$. Analytical support is also provided.
These findings shed new light on the dynamics of blood flow.
\end{abstract}


\maketitle

\paragraph{Introduction }

After nearly a century of research on blood, understanding of the
basic blood flow mechanisms at the cellular level (at the scale of
red blood cells, platelets, etc.) is still an open issue.
Blood is
a complex fluid and the description of its flow properties escapes
the traditional Navier-Stokes law known for simple fluids (e.g. water).

To date blood flow has been described by means of phenomenological
continuum models that require many assumptions which are difficult
both to justify and to validate \cite{Fung}. Furthermore, the modern
view of complex fluids has highlighted the intimate coupling between
the dynamics of the microscopic suspended entities (RBCs in the example
of blood) and the flow at the global scale \cite{Larson}. This implies
that the macroscopic law of blood flow should carry information on
the microscopic scale, such as the orientation of the cells, their
individual and collective dynamics, their local concentration (hematocrit),
and so on.
Apart from a dilute suspension where a rheological law  can be extracted analytically \cite{Danker2007a},
a complete understanding of blood flow should ultimately emerge from
a numerical study.

Computational approaches are, however, challenging due to the free
boundary character of the RBCs (the shape is not known a priori) which
is fixed via a subtle interplay between the local flow and the different
internal modes of the RBCs (membrane bending, shear elasticity).

Under simple shear flow, RBCs exhibit \textit{tank-treading} (TT),
\textit{tumbling} (TB), \textit{vacillating-breathing} (VB) (aka
swinging or trembling) and \textit{{ kayaking}} (K) (see movies of the
four modes \cite{supplement}). These dynamics should have an impact on the
macroscopic flow. These {modes are shared both by vesicles
\cite{Keller1982,Kraus1996,Haas1997,Seifert1999,Kantsler2005,Misbah2006,Mader2006,Vlahovska2007,Danker2007,Lebedev2007,Noguchi2007,Danker2007a}
(made of a pure bilayer phospholipid membrane) and RBCs
\cite{Schonbein1978}, which, apart from having the same type of
phospholipid membrane, are endowed with a cytoskeleton (a
cross-linked network of proteins lying underneath the RBC membrane).
A systematic analysis of these dynamics constitutes an essential step
that is explored in this Letter.

We consider a model system, which is a phospholipid vesicle (Figure \ref{fig:Snapshot}). We thus ignore in this first exploration the effect of the cytoskeleton. This reduction
is essential to identify the basic underlying mechanisms.
.

\begin{figure}
\includegraphics[width=0.8\columnwidth]{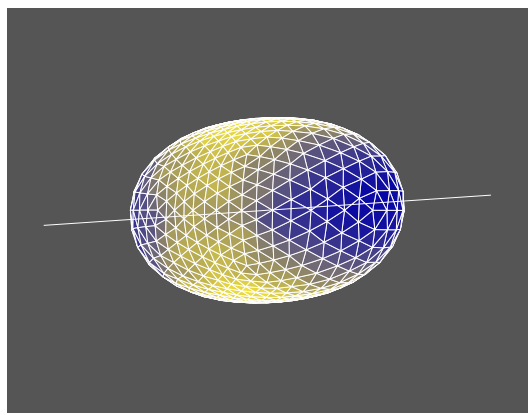}
\caption{\label{fig:Snapshot}A typical vesicle obtained by simulations. The white line indicates the main axis direction.
The color code on the vesicle corresponds to the value of the local
curvature, which is large at the tips and low in the central region.}
\end{figure}

The major results of this study are:  (i) We find that the
width of the region of parameters where the VB mode  is manifested significantly widens upon increasing shear rate. This behavior was
not captured by any of the previous theories or simulations. (ii) We
find that the period of oscillation in the TT and TB regime is
practically independent of the shear rate $\dot\gamma$, in marked
contrast with a previous numerical study using multiparticle
collision dynamics \cite{Noguchi2007}. (iii) We develop an
analytical theory that accounts for these two facts. (iv) We demonstrate
that the Keller-Skalak theory (which assumes a fixed shape for the
RBCs, and is often used as a
basis in experiments and theoretical studies) is an excellent framework for TT and TB description when
shape deformability is negligible (small $\dot\gamma$ or large
$\lambda$, the ratio of internal over the external viscosity).
However, major deviations are revealed when $\dot\gamma$ is large
enough and/or $\lambda$ is not too large. {(v) We report on a
transition between tumbling and the kayaking phase, during which the main
axis of the vesicle rotates around an axis perpendicular to the
shear plane. Strikingly, kayaking seems
 to occur even if the vesicle is initially forced to be in the shear plane.}

\paragraph{The Model:}

The Stokes equations (experiments have so far explored the limit of
small Reynolds numbers) can be formally solved using the Boundary
Integral (BI) formalism (see for example \cite{Pozrikidis92}) which
yields

\begin{eqnarray}
\begin{array}{cc}
 & \eta_{m}\mathbf{v}_{mem}(\mathbf{r})=\eta_{out}\mathbf{v}_{shear}+\int_{mem}\overline{G}(\mathbf{r}-\mathbf{r}^{'})\mathbf{f}_{mem}(\mathbf{r}^{'})\: d\mathbf{r}^{'}\\
 & +(\eta_{in}-\eta_{out})\int_{mem}\mathbf{v}_{mem}(\mathbf{r}^{'}).\overline{K}(\mathbf{r}-\mathbf{r}^{'}).\hat{\mathbf{n}}(\mathbf{r}^{'})\: d\mathbf{r}^{'}\end{array}\label{eq:integral}\end{eqnarray}
 where $\mathbf{v}_{mem}(\mathbf{r})$ is the local velocity field
at the membrane, $\mathbf{v}_{shear}=\dot\gamma y {\bf\hat x}$ is
the externally applied Couette flow (with $\dot\gamma$ the shear
rate), $\eta_{m}\equiv (\eta_{in}+\eta_{out})/2$ ($\eta_{in}$ and
$\eta_{ou}$ stand for viscosities of the internal and external
fluid, respectively), $\int_{mem}$ is an integration over the
membrane and $\overline{G}(\mathbf{r}-\mathbf{r}^{'})$ and
$\overline{K}(\mathbf{r}-\mathbf{r}^{'})$ are the Green tensors
defined as:\[
\overline{G}(\mathbf{x})_{ij}=\frac{1}{8\pi}\left(\frac{\delta_{ij}}{x}+\frac{x_{i}x_{j}}{x^{3}}\right);
\overline{K}(\mathbf{x})_{ijk}=\frac{3}{4\pi}\frac{x_{i}x_{j}x_{k}}{x^{5}}{\normalcolor
}\] A vesicle that is subjected to a flow undergoes a shape
transformation that is limited by bending modes. The reaction
bending force of a vesicle on the fluid is given by the Helfrich
force \cite{Helfrich73}
\begin{equation}
\mathbf{f}_{curv}(\mathbf{r})=-\kappa\left[\frac{1}{2}c(\mathbf{r})\left\{ c(\mathbf{r})^{2}-4g(\mathbf{r})\right\} +\Delta_{2D}c(\mathbf{r})\right]\hat{\mathbf{n}}(\mathbf{r})
\end{equation}
where $\kappa$ is the bending modulus, $c(\mathbf{r})$ is the local
mean curvature ($c(\mathbf{r})=c_{1}(\mathbf{r})+c_{2}(\mathbf{r})$,
$c_{1}$ and $c_{2}$ are the two principal curvatures at point
$\mathbf{r}$ of the membrane), $g(\mathbf{r})$ is the Gaussian
curvature ($g=c_{1}c_{2}$), $\Delta_{2D}$ is the (surface)
Laplace-Beltrami operator on the membrane, and $\hat{\mathbf{n}}$ is
the outward unit normal vector. The other contribution to the
membrane force follows from
local membrane incompressibility:\[
\mathbf{f}_{tens}(\mathbf{r})=T\left[\zeta(\mathbf{r})c(\mathbf{r})\hat{\mathbf{n}}(\mathbf{r})+\mathbf{\nabla}_{2D}\zeta(\mathbf{r})\right]\]
where $\mathbf{\nabla}_{2D}$ is the surface gradient operator and
$\zeta(\mathbf{r})$ is a local dimensionless Lagrange multiplier
that enforces membrane inextensibility (the constant $T$ has a
dimension of  energy per unit surface).
Local membrane inextensibility sets severe limitations on the
numerics, both on the time step and the precision of the results.
Carelessness (for example, imposing a small enough tolerance of a few
$\%$ on area) may induce spurious results. The scheme used in this
study ensures an area variation lower than $5.10^{-4}\%$.
The precise numerical analysis is a formidable task and technical details will be reported elsewhere. Instead, we focus here on the major outcomes.

\paragraph{Results:}

The  phase diagram of  various dynamics (TT, TB and VB) of a vesicle and RBC in a Couette flow recently motivated
a large number of theoretical  and experimental studies \cite{Noguchi2007,Danker2007a,Lebedev2007,Deschamps2009,Secomb2007}. This is the first focus of our study.
Let us introduce appropriate dimensionless numbers. Shape
deformability can be measured by the dimensionless number

\[
C_{\kappa}=\frac{\eta_{out}\gamma R^{3}}{\kappa}\]
where $R$ is the typical size of the vesicle.
Vesicles exhibit various equilibrium shapes (i.e., in the absence of
flow) \cite{Seifert1997}, depending on their reduced volume
$\tau=6\sqrt{\pi}V/S^{3/2}$, where $S$ is the area of the membrane
and $V$ the internal volume. $\tau$ is the second dimensionless
parameter. For $\tau>0.652$, the shape is prolate  (one long
revolution axis) \cite{Seifert1997}. In the range $0.591
<\tau<0.652$, the shape is oblate (one small revolution axis; the
shape is biconcave, known also for RBCs). We found that, for
$C_{\kappa}\geq1$ (which is quite easily reachable experimentally),
the oblate branch is suppressed in a Couette flow so that only
prolate shapes prevail (see Fig.\ref{fig:Snapshot}).
The last dimensionless parameter is  $\lambda=\eta_{in}/\eta_{out}$.
While our study can be performed at different $\tau$, we will
consider a reduced volume close enough to unity. This will allow us to
compare the theoretical approaches considering the
quasi-spherical limit
\cite{Misbah2006,Vlahovska2007,Danker2007,Lebedev2007,Noguchi2007}.
We have chosen the value $\tau=0.95$ and have explored the effects
of the two other parameters $C_\kappa$ and $\lambda$. The resulting
phase diagram is represented in Fig.\ref{fig:Phase-diagram}.

 A first important feature found here is that the boundaries of the phase diagram
are strongly underestimated in previous analytical theories
\cite{Danker2007,Lebedev2007}. For example, for $C_\kappa\simeq 0.1$,
the bifurcation from TT to TB occurs here for $\lambda\simeq 8$,
while analytical theories predicted $\lambda \simeq 4$
(Fig.\ref{fig:Phase-diagram}). This is astonishing because
deviation from a sphere is only about $5\%$ and a perturbative
scheme is expected to make sense. Thus, we have attempted to
understand this peculiar behavior by revising previous analytical
theories. The key ingredient is that the next order terms in an
expansion in powers of excess area (relative to a sphere)
 are decisive, however small the deviation from a sphere.
 We have identified the fact that this is the result of a singular behavior of the expansion scheme.
  Details of the analytical theory will be given elsewhere \cite{Farutin2009}.
  The outcome of this analytical theory is presented in Figure \ref{fig:Phase-diagram},
   revealing excellent agreement with the full numerical simulation.

\begin{figure}
\includegraphics[width=1\columnwidth]{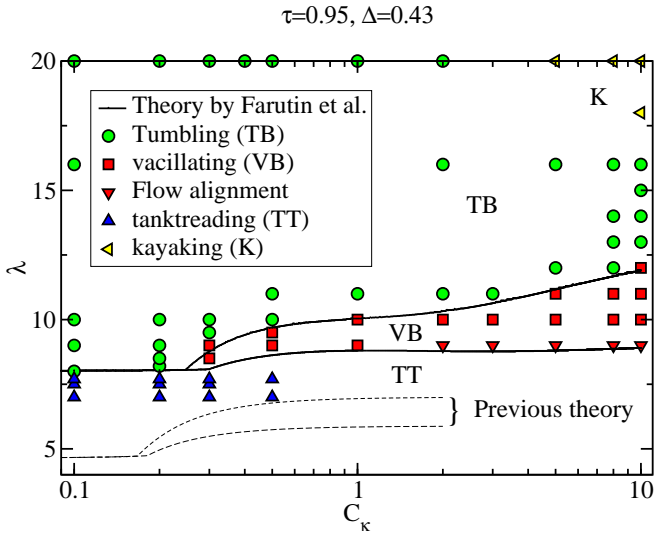}
\caption{\label{fig:Phase-diagram}Phase diagram for $\tau=0.95$. The symbols
 correspond to numerical results and the solid line to the analytical. The Flow alignment states correspond
to a VB mode with a very small amplitude  (see Fig.\ref{fig:relax}). These states are thus
intermediate between VB and TT. {The K phase is visible at the upper right corner of the diagram.}}
\end{figure}

Previous analytical \cite{Lebedev2007,Danker2007a} as well as numerical calculations \cite{Noguchi2007} (based on dissipative particle dynamics) and experiments \cite{Deschamps2009} reported that the band of existence of the VB mode saturates with $C_\kappa$ above a value of the order of $C_\kappa\sim 0.5$. This is in good agreement with the present numerical study as long as $0.5<C_\kappa <2$. Beyond a value of the order of $C_\kappa=2$, the VB band exhibits a sudden ample widening as shown in Fig. \ref{fig:Phase-diagram}. This effect highlights the nontrivial character of dynamics due to shape deformation.
A careful theoretical analysis of this phenomenon has led us to the discovery that not only is the fourth order harmonic strongly coupled to the second one, but it also acquires strong activity on increasing $C_\kappa$.
This mode, which is damped at low deformability,
becomes active (i.e. it is excited) at larger deformability (larger
$C_\kappa$). As a consequence, the VB regime is promoted further and
further, leading to the VB band widening. We have found that this
peculiar behavior is captured by the new theory that implements
fourth order harmonics \cite{Farutin2009}.

Let us now analyze the behavior of the tumbling angle $\theta (t)$. It is convenient to represent a phase portrait in the plane $(\dot{\theta},\theta )$. This will also allow us to shed light on the limit of applicability of the KS theory, which is often used as a basis in experiments \cite{Mader2006,Kantsler2005} and in numerical simulations \cite{Noguchi2007}.
We find that, for $C_{\kappa}\to0$ (we choose $C_\kappa=0.1$ as an example), the simple relation $\dot{\theta}/\gamma=A+B\cos(2\theta)$
predicted by the KS theory is in excellent agreement with the BI simulations.

\begin{figure}
\includegraphics[width=1\columnwidth]{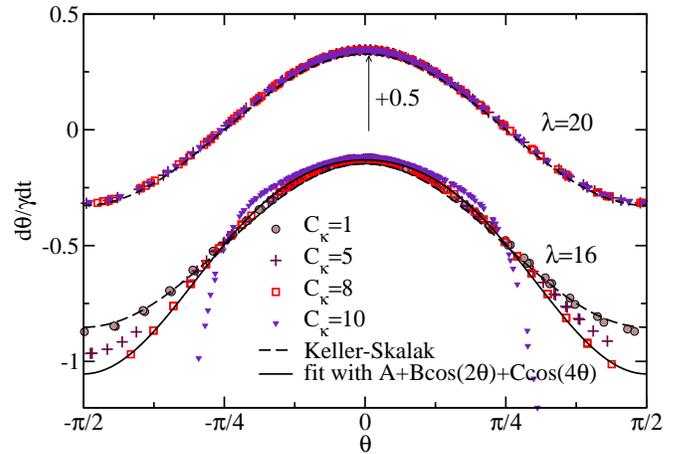}
\caption{\label{fig:KS2}$\dot{\theta}$ as a function of $\theta$ for
$\lambda=16$, $\lambda=20$ (shifted upwards by +0.5),  for various
$C_{\kappa}$. The dashed line corresponds to the KS theory. The full
black line is a fit of the data points for$\lambda=16$ and
$C_{\kappa}=8$ with the function $A+B\cos(2\theta)+C\cos(4\theta)$
($A=0.543$, $B=0.462$ and $C=0.049$), showing that higher harmonics
play an important role at large deformabilities. }
\end{figure}

At larger deformabilities, the situation is more complex and requires
a careful analysis. We first focus on the value  $\lambda=20$. The
result is reported in Fig.\ref{fig:KS2} (shifted by $+0.5$ upward
for  clarity). All data for $1<C_{\kappa}<10$ are nearly
superimposed, reflecting a good agreement with the KS theory {even
though, for $C_{\kappa}\geq5$, the vesicle is actually kayaking \cite{Lebedev2007}.
$\theta(t)$ represents in this case the tumbling of the projection
of the main axis in the shear plane.} In contrast, at smaller
values of $\lambda$, ($\lambda=16$ in Fig.\ref{fig:KS2}), major
deviations from the KS theory are manifested. In particular, we
observe the excitation of the fourth order harmonic,
 represented by $\cos(4\theta)$ in the figure when $C_{k}$ is increased
up to $8$. Note that, for  $C_{\kappa}=10$ ,other higher order
harmonics are excited as well, which is a precursor to the TB-VB  {
or the TB-K} bifurcation.

 We have analyzed several other physical quantities, such as evolution and possible exchanges of the two main axes lying in the shear plane, but we shall focus only on the behavior of the period (scaled by the shear rate) in this brief exploration. The results are shown in
 Fig.\ref{fig:period}. A remarkable fact is the quasi-robust character of the rescaled period as a function of the deformation {regardless of
  the dynamical mode (TB, VB or K)} (only minor variations by few $\%$ are manifested over a range of a decade in shear rate). This finding is
  not intuitive in as much as deformation (at large enough $C_\kappa$) plays an essential role (strong deviation from  the KS theory). This result runs contrary to a previous numerical analysis \cite{Noguchi2007} based on dissipative particle dynamics. The ratio of their period variation and ours attains a factor $100$. We do not have an explanation about the origin of this major discrepancy. We have checked, in light of the new analytical theory, that the same quasi-independence on $C_\kappa$ is also manifested.

\begin{figure}
\includegraphics[width=1\columnwidth]{period_tau=0\lyxdot 95}
\caption{\label{fig:period} Rescaled tumbling period for $\tau=0.95$ as a function
of the deformability $C_{\kappa}$ for $\lambda=11$, $16$ and $20$.}
\end{figure}
\begin{figure}
\includegraphics[width=1\columnwidth]{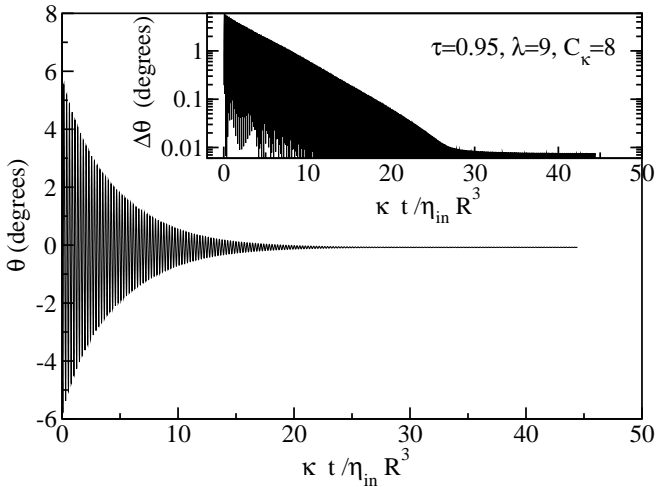}
\caption{\label{fig:relax}Long time relaxation of the vesicle close to the
TT-VB line. The insert shows a log plot of the vacillation amplitude.}
\end{figure}

Finally, it is worth emphasizing that the transient dynamical regime can prove to be very slow,
 especially at large values of $C_{\kappa}$. It is necessary
to ascertain the establishment of a given regime on the scale of the
longest time. For large enough $\lambda$, the appropriate (slow)
time scale is $\tau_{\kappa}= \eta_{in} R^3/\kappa$; motion is
limited by the more viscous internal fluid. For $C_{\kappa}=10$,
this may typically represent a hundred cycles in the TB
regime. Thus, it is necessary to make use of appropriate and stable
numerical schemes. Details will be published elsewhere. Here, it suffices to provide an illustration exhibiting the slow relaxation
(Fig.\ref{fig:relax}). The VB amplitude close to the TT-VB line
($\lambda=9$, $C_{\kappa}=8$) is shown. We find an
 exponential relaxation of the VB amplitude (the insert
of Fig.\ref{fig:relax} provides a log plot of the amplitude). The
vesicle reaches its steady state only after a characteristic time of about
$30\eta_{in}R^{3}/\kappa$, corresponding to 125 oscillations. The same feature is observed for the TB-K transition.%

\noindent{\it Conclusion:} We have reported on a quantitative
numerical analysis in three dimensions that led us to identify new features of vesicle dynamics, which were not revealed in prior analytical, numerical, or
experimental studies. A theory has allowed us to unearth
a subtle role played by higher order terms. The theoretical results
agree remarkably well with the numerical ones.
This work is a first essential step toward a
systematic study of RBC dynamics and blood rheology. An obvious
limitation of our work is the absence of shear elasticity associated
with the cytoskeleton. An analysis that includes this factor constitutes
the next natural step.

C.M. and A.F. Acknowledge financial support from CNES and ANR (MOSICOB project).

\end{document}